\documentclass[12pt]{iopart}
\usepackage{graphicx}
\begin{document}
\title[Exponential Operators, Dobi\'nski Relations and Summability]{Exponential Operators, Dobi\'nski Relations and Summability}
\author{P. Blasiak$^{1,2}$, A. Gawron$^{1}$, A. Horzela$^{1}$, K. A. Penson$^{2}$ 
\\
and A. I. Solomon $^{2,3}$}
\address{$^1$ H.Niewodnicza\'nski Institute of Nuclear Physics, Polish Academy of
Sciences,\\ ul. Eliasza-Radzikowskiego 152, PL 31342 Krak\'ow, Poland}
\address{$^2$Laboratoire de Physique Th\'eorique de la Mati\`ere
Condens\'ee,
Universit\'e P.~et~M.~Curie, Tour 24 - 2e \'et., 4 Pl.Jussieu, F 75252 Paris Cedex 05, France}
\address{$^3$Physics and Astronomy Department, The Open University,\\
Milton Keynes MK7 6AA, United Kingdom}
\ead{pawel.blasiak@ifj.edu.pl,agnieszka.gawron@ifj.edu.pl,
\\
andrzej.horzela@ifj.edu.pl, penson@lptl.jussieu.fr, 
\\
a.i.solomon@open.ac.uk}
\begin{abstract}
We investigate properties of exponential operators preserving the particle number using combinatorial methods developed in order to solve the boson normal ordering problem. In particular, we apply generalized Dobi\'nski relations and methods of multivariate Bell polynomials which enable us to understand the meaning of perturbation-like expansions of exponential operators. Such expansions, obtained as formal power series, are everywhere divergent but the Pad\'e summation method is shown to give results which very well agree with exact solutions got for simplified quantum models of the one mode bosonic systems.   
\end{abstract}
\pacs{03.65.Fd,~02.10.Ox,~02.30.Lt}
\maketitle
Consider exponential operators of the form $\exp{(-\lambda{\cal H}_{\boldmath{\alpha}}({\hat n}))}$, where 
${\cal H}_{\boldmath{\alpha}}({\hat n})= \sum_{i=1}^{N}\alpha_{i}{\hat n}^{i}$ is a polynomial of the number operator ${\hat n}$ (assumed to be bounded from below), $\alpha_{i}$'s play roles of coupling constants and $\lambda$ is an overall positive parameter. Calculation of the number state representation of $\exp{(-\lambda{\cal H}_{\boldmath{\alpha}}({\hat n}))}$  is straightforward
\begin{equation}
\label{1}
\langle l|\exp{(-\lambda{\cal H}_{\boldmath{\alpha}}({\hat n}))}|m\rangle = \exp{(-\lambda\sum_{i=1}^{N}\alpha_{i}{m}^{i})}\delta_{lm},
\end{equation}
\noindent as well as its (standard) coherent state representation ($|z\rangle = 
\exp{(-\frac{|z|^2}{2})}\sum_{n=0}^{\infty}\frac{z^n}{\sqrt{n!}}|n\rangle$, ${\hat n}|n\rangle = n|n\rangle$, 
$\langle n|n^{\prime}\rangle = \delta_{nn^{\prime}}$, 
$a|z\rangle = z|z\rangle$ and $\langle z^{\prime}|z\rangle = \exp{\left(-\frac{1}{2}\left(|z^{\prime}|^2 + |z|^{2} - 2z^{\prime}z\right)\right)}$)
\begin{equation}
\label{2}
\langle z^{\prime}|\exp{(-\lambda{\cal H}_{\boldmath{\alpha}}({\hat n}))}|z\rangle = \langle z^{\prime}|z\rangle e^{-z^{\prime}z}
\sum\limits_{k=0}^{\infty}\frac{\displaystyle (z^{\prime}z)^{k}}{\displaystyle k!}\exp{\left(-\lambda\sum_{i=1}^{N}\alpha_{i}{k}^{i}\right)}.
\end{equation}
\noindent For the simplest example of ${\cal H}_{\boldmath{\alpha}}({\hat n})= {\hat n}$ the r.h.s of Eqn.(\ref{2}) is given in terms of the elementary function
\begin{equation}
\label{3}
\begin{array}{rcl}
\langle z^{\prime}|\exp{(-\lambda{\hat n})}|z\rangle &=& \langle z^{\prime}|z\rangle e^{-z^{\prime}z}\sum\limits_{k=0}^{\infty}\frac{\displaystyle (z^{\prime}z)^{k}}{\displaystyle k!}\exp{(-\lambda k)}\\ 
&=& \langle z^{\prime}|z\rangle\exp{\left(z^{\prime}z\left(e^{-\lambda} - 1\right)\right)}
\end{array}
\end{equation}
\noindent in which one recognizes the exponential generating function of the (exponential) Bell polynomials \cite{katriel}
\begin{equation}
\label{4}
\exp{\left(x\left(e^\lambda - 1\right)\right)} = \sum\limits_{k=0}^{\infty}\frac{\displaystyle \lambda^{k}}{\displaystyle k!} B(k,x).
\end{equation}
\noindent The Bell polynomials are well known from their applications in combinatorics \cite{comtet}. They are defined as
\begin{equation}
\label{5}
B(n,x) = \sum\limits_{k=1}^n S(n,k) x^k, 
\end{equation}
\noindent where $S(n,k)$ denote the \emph{Stirling numbers of the second kind} (positive integers which in enumerative combinatorics count the number of ways of putting $n$ different objects into $k$ identical containers leaving none container empty) whose analytic representation is
\begin{equation}
\label{6}
S(n,k) = \frac{1}{k!}\sum\limits_{j=1}^k\left(\begin{array}{c}{k}\\{j}\end{array}\right)(-1)^{k-j}j^n.
\end{equation}
\noindent Particular values of the Bell polynomials $B(n) = B(n,1)$ are known as the Bell numbers and in enumerative combinatorics count the number of ways of putting $n$ different objects into $n$ identical containers some of which may be left empty. This means that the Bell numbers give us the number of partitions of an $n$-element set. Expanding the l.h.s. of the Eqn.(\ref{3}) as a power series in $\lambda$ and using the definition of the coherent states we arrive at 
\begin{equation}
\label{7}
\langle z^{\prime}|z\rangle e^{-z^{\prime}z}\sum\limits_{k=0}^{\infty}\frac{\displaystyle \lambda^{k}}{\displaystyle k!}
\sum\limits_{m=0}^{\infty}\frac{\displaystyle (z^{\prime}z)^{m}}{\displaystyle m!}
m^{k} = \langle z^{\prime}|z\rangle
\sum\limits_{k=0}^{\infty}\frac{\displaystyle \lambda^{k}}{\displaystyle k!} B(k,z^{\prime}z) 
\end{equation}
\noindent from which one reads out the formula
\begin{equation}
\label{8}
e^{-z^{\prime}z}\sum\limits_{m=0}^{\infty}\frac{\displaystyle (z^{\prime}z)^{m}}{\displaystyle m!}
m^{k} = 
B(k,z^{\prime}z), 
\end{equation}
\noindent giving for $z^{\prime}=z=1$ the Dobi\'nski relation
\begin{equation}
\label{9}
B(n) = e^{-1}\sum\limits_{m=0}^{\infty}\frac{\displaystyle m^{k}}{\displaystyle m!}.
\end{equation}
\noindent Eqns. ({\ref 8}) and (\ref{9}) connect sequences of polynomials $B(k,z^{\prime}z)$ or positive integers $B(n)$ with sums of nontrivial series of fractions and allow to represent the Bell polynomials as the Stieltjes moments of an infinite sum of weighted $\delta$-functions, called the {\it Dirac comb}
\begin{equation}
\label{10}
B(n,x) = e^{-x}\int_{0}^{\infty}{\rm d}y y^{n}\sum\limits_{m=0}^{\infty}\frac{\displaystyle x^{m}}{\displaystyle m!}\delta(y-m).
\end{equation}
\noindent Following the above considerations  we can generalize our results  to arbitrary $\exp{(-\lambda{\cal H}_{\boldmath\alpha}({\hat n}))}$ in which ${\cal H}_{\boldmath\alpha}({\hat n})$ polynomially depend on ${\hat n}$. For such a case we get
\begin{equation}
\label{11}
\langle z^{\prime}|\exp{(-\lambda{\cal H}_{\boldmath\alpha}({\hat n}))}|z\rangle =\langle z^{\prime}|z\rangle\sum_{k=0}^\infty
B_{\boldmath{\alpha}}(k,z^{\prime}z)\frac{(-\lambda)^k}{k!},
\end{equation}
\noindent where generalized Bell polynomials $B_{\boldmath{\alpha}}(m,z^{\prime}z)$ are defined through generalized Stirling numbers of the second kind \cite{blasiak1}, \cite{blasiak2}, \cite{blasiak3} 
\begin{equation}
\label{12}
\begin {array}{c}
B_{\boldmath{\alpha}}(m,x)=\sum\limits_{k=1}^{mN}S_{\boldmath{\alpha}}(n,k)\
x^k,
\\
S_{{\boldmath\alpha}}(n,k)=\frac{\displaystyle 1}{\displaystyle k!}
\sum\limits_{j=0}^k\left(\begin{array}{c}{k}\\{j}\end{array}\right)(-1)^{k-j}
\left[\sum\limits_{l=1}^N{\bar \alpha}_l\
j^{\underline{l}}\right]^n,
\end{array}
\end{equation}
\noindent with $j^{\underline{l}} = j(j-1)\dots(j-l+1)$ denoting the falling factorial and ${\bar \alpha}_l=\sum_{m=1}^ls(l,m)\alpha_{m}$ being the inverse Stirling transform of $\alpha_i$ given by the Stirling numbers of the first kind $s(n,k)$, $\sum_{k=m}^ns(n,k)S(k,m)=\delta_{nm}$. Generalized Dobi\'nski relations read now \cite{blasiak2}, \cite{blasiak3}
\begin{equation}
\label{13}
B_{{\boldmath\alpha}}(n,x)=e^{-x}\sum\limits_{m=0}^\infty\left[
\sum\limits_{k=1}^N{\bar\alpha}_k\
m^{\underline{k}}\right]^n\frac{x^m}{m!}
\end{equation}
\noindent and, analogously to Eqn.(\ref{10}), provide us with representation of $B_{{\boldmath\alpha}}(n,x)$ as moments
\begin{equation}
\label{14}
B_{{\boldmath\alpha}}(n,x)= e^{-x}\int_{\Delta}{\rm d}y y^{n}\sum\limits_{m=0}^{\infty}\frac{\displaystyle x^{m}}{\displaystyle m!}\delta\left(y-\sum\limits_{k=1}^N{\bar\alpha}_k\
m^{\underline{k}}\right),
\end{equation}
\noindent where the domain of integration $\Delta=[\,{\inf\left(\sum_{k=1}^N{\bar\alpha}_k\
m^{\underline{k}}\right)-\varepsilon},{\infty})$. Eqn.(\ref{14}), if put into Eqn.(\ref{11}) and changed the summation order, leads to Eqn.(\ref{2}). Note that using the generalized Dobi\'nski formula we give analytical meaning to the formal series (\ref{11}). As a rule these series are divergent because the coefficients $B_{{\boldmath\alpha}}(n,x)$ grow with $n$ much faster than $n!$. Such an asymptotic behavior is seen from Eqns.(\ref{12}) - the latter imply that the numbers $S_{{\boldmath\alpha}}(n,k)$ include the standard Stirling numbers of the second kind $S(nN,k)$ and, as a consequence, the polynomials $B_{{\boldmath\alpha}}(n,x)$ include polynomials $B(nN,x)$. The $n\rightarrow\infty$ asymptotics of the standard Bell numbers $B(n,1)$ is $B_n\sim n!\frac{\exp\bigl(\exp{(r(n))}-1\bigr)}
{[r(n)]^{n+1}\sqrt{2\pi\exp{(r(n))}}}$ where $r(n)\sim \log{n}-\log(\log{n})$ and it causes that the series (\ref{11}) are divergent for $N\ge 2$.

For the toy model ${\cal H}_{\boldmath{\alpha}}({\hat n})= {\hat n}$ we were able to find the closed form of $\langle z^{\prime}|\exp{(-\lambda{\hat n})}|z\rangle$ given in terms of elementary functions  - {\it i.e.} we solved explicitly the normal ordering problem for such an operator \cite{louisell}. If ${\cal H}_{\boldmath{\alpha}}({\hat n})$ becomes a more complicated polynomial then the problem complicates but it remains manageable and gives some insight into perturbation methods widely used in quantum mechanics and quantum field theory. Because in the following we are going to concentrate ourselves on the problems related to the coupling constant perturbation calculus treated with combinatorics-based methods we do not use the scheme leading to the generalized Bell polynomials but we will investigate the problem using methods of multivariate Bell polynomials, \cite{comtet}, \cite{aldrovandi}, still emphasizing the importance of the Dobi\'nski-type relations.  The multivariate Bell polynomials enable us to 
construct the Taylor--Maclaurin expansion of a composite
function $f(g(x))$. To this end let us recall that for any $f(x)=\sum_{n=1}^\infty
f_nx^n/n!$ and $g(x)=\sum_{n=1}^\infty
g_nx^n/n!$ given as formal power series one gets
\begin{equation}
\label{15} f(g(x))=[f\circ g](x)=\sum_{n=1}^\infty
\left(\sum_{k=1}^nB_{nk}(g_1,g_2,\dots,g_{n-k+1})f_k\right)\,\frac{x^n}{n!}\,,
\end{equation}
\noindent where the coefficients $B_{nk}$ are certain polynomials in the Taylor
coefficients $g_i$ - namely the multivariate Bell polynomials - given by
\begin{equation}
\label{16} {B}_{n,k}(g_1,\dots,g_{n-k+1})=
{\sum_{\{\nu_i\}}}^{\prime\prime}\frac{n!}
{\prod\limits_{j=1}^{n}\left[\nu_j!\,(j!)^{\nu_j}\right]}\,
g_1^{\nu_1}g_2^{\nu_2}\dots g_{n-k+1}^{\nu_{n-k+1}}\,,
\end{equation}
\noindent where the summation ${\sum\limits_{\{\nu_i\}}}^{\prime\prime}$ is
over all possible non-negative $\{\nu_i\}$ being partitions of
an integer $n$ into sum of $k$ integers, {\it i.e.} over $\{\nu_i\}$ being solutions to 
the equations $\sum_{j=1}^n j\nu_j=n$ and $\sum_{j=1}^n\nu_j=k$\footnote{ This condition shows that the multivariate Bell polynomials are closely related to combinatorial numbers}. 
Eqns.(\ref{15}) and (\ref{16}) imply that the
multivariate Bell polynomials satisfy, for $a$ and $b$ arbitrary
constants, the homogeneity relation
\begin{equation}
\label{18} B_{n,k}(ab^1g_1,ab^2g_2,\dots,ab^{n-k+1}g_{n-k+1})
=a^k b^n\,B_{n,k}(g_1,g_2...,g_{n-k+1})\,.
\end{equation}
\noindent A particular case of the multivariate Bell polynomials are polynomials being coefficients of Taylor-Maclaurin expansions of $\exp(\sum_{i=0}^{N}\alpha_{i}x^{i})$. They generalize the standard Hermite polynomials and for some special cases have simple analytic forms \cite{dattoli1}, \cite{dattoli2} - among them the two variable
Hermite--Kamp\'e de F\'eriet polynomials $H_n^{(M)}(x,y)$:   
\begin{equation}
\label{19}
\begin{array}{c}
\displaystyle\exp\left(g_1x+g_M\,{x^M}\right)=
\sum_{n=0}^{\infty}H_n^{(M)}(g_1,{g_{M}})\frac{\displaystyle x^n}{\displaystyle n!}\,,\\[6pt]
H_n^{(M)}(g_1,{g_{M}})=n!\displaystyle\sum_{r=0}^{[n/M]}
\frac{g_1^{n-Mr}g_{M}^r}{(n-Mr)!r!}\,,
\end{array}
\end{equation}
\noindent  and the three-variable Hermite polynomials $H_n(a,b,c)$:
\begin{equation}
\label{20}
\begin{array}{c}
\displaystyle\exp\left(ax+bx^2+cx^3\right)=
\sum_{n=0}^{\infty}H_n(a,b,c)\frac{x^n}{n!}\,,\\[6pt]
H_n(a,b,c)=n!\displaystyle\sum_{r=0}^{[n/3]}
\frac{c^{r}H_{n-3r}^{2}(a,b)}{(n-3r)!r!}\,.
\end{array}
\end{equation}
 
As an illustration of the presented approach let us consider diagonal coherent state matrix element $\langle z|\exp{\left(-\left(g{(\xi\hat n)} + G{(\xi\hat n})^2\right)\right)}|z\rangle$. Expanding the exponential as power series in ${\xi}$, next using (\ref{18}), (\ref{19}) and definition of the coherent states we arrive at 
\begin{equation}
\label{20}
\langle z|\exp{\left(-\left(g{(\xi\hat n)} + G{(\xi\hat n)}^2\right)\right)}|z\rangle = 1 + \sum\limits_{n=1}^{\infty}H_{n}^{(2)}(-g,-G)B({n,|z|^2})\frac{\displaystyle \xi^n}{\displaystyle n!}.
\end{equation}
\noindent The operational relation
$H_{n}^{(2)}\left(x,y\right) = \exp{\left(y\frac{\partial^2}{\partial {x}^2}\right)}x^n$
enables us to rewrite the Eqn.(\ref{20}) as
\begin{equation}
\label{22}
\begin{array}{rcl}
\langle z|e^{-\left(g{(\xi\hat n)} + G{(\xi\hat n)}^2\right)}|z\rangle 
&=&
\exp{\left(-G\frac{\displaystyle\partial^2}{\displaystyle\partial { g}^2}\right)}\exp{\left(|z|^2\left(e^{-g\xi}-1\right)\right)}\\
&=&
\sum\limits_{k=0}^{\infty}\frac{\displaystyle (-G)^k}{\displaystyle k!}\sum\limits_{l=0}^{\infty}B(l,|z|^2)\,\frac{\displaystyle\xi^{l}}{\displaystyle l!}\,\frac{\displaystyle{\rm d}^k}{\displaystyle {\rm d}{g}^k}\,(-g)^l\\
&=&
\sum\limits_{k=0}^{\infty}(-G)^k\frac{\displaystyle\xi^{2k}}{\displaystyle k!}\sum\limits_{i=0}^{\infty}B(2k+i,|z|^2)\frac{\displaystyle (-g\xi)^i}{\displaystyle i!}.
\end{array}
\end{equation}
\noindent Replacing $B(2k+i,|z|^2)$ by their moment representation (\ref{10}) and changing the integration and summation order we obtain the perturbation expansion of $\langle z|\exp{\left(-\left(g{(\xi\hat n)} + G{(\xi\hat n)}^2\right)\right)}|z\rangle$ in terms of the power series in the coupling constant $G$
\begin{equation}
\label{23}
\begin{array}{rcl}
\langle z|e^{-\left(g{(\xi\hat n)} + G{(\xi\hat n)}^2\right)}|z\rangle 
&=&
e^{|z|^2\left(e^{-g\xi}-1\right)}\sum\limits_{k=0}^{\infty}(-G)^k\frac{\displaystyle \left(\xi^2\right)^{k}}{\displaystyle k!}B(2k,|z|^2e^{-g\xi})
\end{array}
\end{equation}
\noindent Because of the asymptotic behavior $B(2k,|z|^2e^{-g\xi})\sim (2k)!, k\rightarrow\infty$, this series, as well as the series (\ref{20}), both have zero radii of convergence being however asymptotic expansions of 
\begin{equation}
\label{24}
\langle z|\exp{\left(-\left(g{(\xi\hat n)} + G{(\xi\hat n)}^2\right)\right)}|z\rangle =
e^{-|z|^{2}}
\sum\limits_{k=0}^{\infty}\frac{\displaystyle |z|^{2}}{\displaystyle k!}e^{-\left(g(\xi k) + G(\xi k)^2\right)}.
\end{equation}
In order to give them analytical meaning one has to use methods of generalized summation. Numerical check using the Pad\'e method (see Fig.1-2, below) shows that even low order approximants of series (\ref{20}) and (\ref{23}) give very good agreement with exact result calculated from (\ref{24}) for G belonging to the domain much larger than the domain in which partial sums of both series give acceptable results. Moreover, comparing these results we in fact compare results given by the Pad\'e method with exact solutions for simplified, nevertheless essentially quantum, model. This confirms practical utility of the Pad\'e summation method applied to various perturbation expansions occurring in quantum physics even if we are unable to prove its applicability in a mathematically satisfactory way. It also confirms that generating functions obtained as solutions to the boson normal ordering problem and being in general divergent formal series may be interpreted as asymptotic expansions, resumed in such a generalized sense and, as a consequence, used in physical applications. 
\section*{Acknowledgments}
{A.G. wishes to thank the Polish Ministry of Scientific Research and
Information Technology for support under Grant No: 1P03B 060 27.}  
\begin{figure}[ht]
\begin{center}
\vspace{0.5mm}
\includegraphics*[angle=0,width=0.42\textwidth]{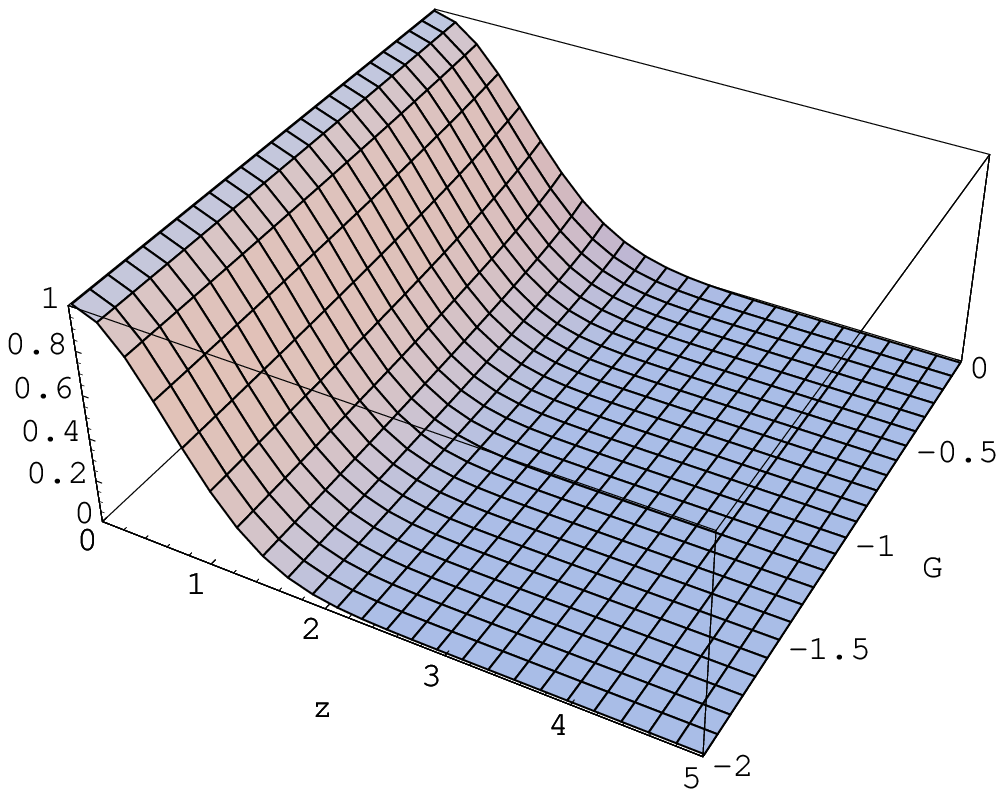}
\hspace{5.mm}
\includegraphics*[angle=0,width=0.47\textwidth]{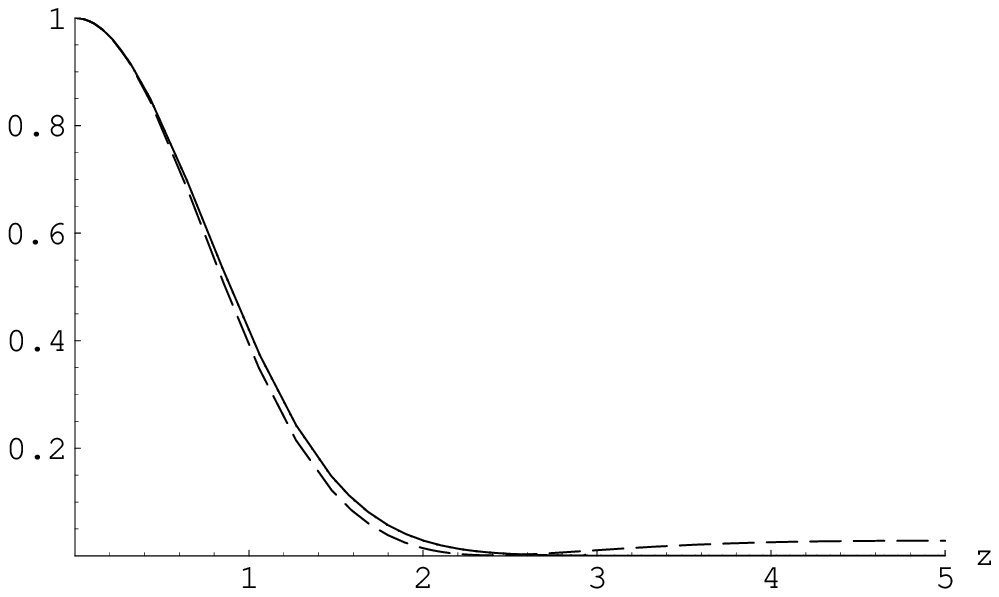}
\end{center}
\vspace{0.5mm}\caption{The sum of the series (\ref{24}), $g=\xi=1$, (left plot) and its comparison with Pad\'e approximants $[3,4]$ for the series (\ref{20}) (dashed curve, worse approximation) and (\ref{23}) (continuous curve, nondistinguishable from the plot of (\ref{24})) for $G=1$. }
\end{figure}
\begin{figure}[ht]
\vspace{0.5mm}
\begin{center}
\includegraphics*[angle=0,width=0.41\textwidth]{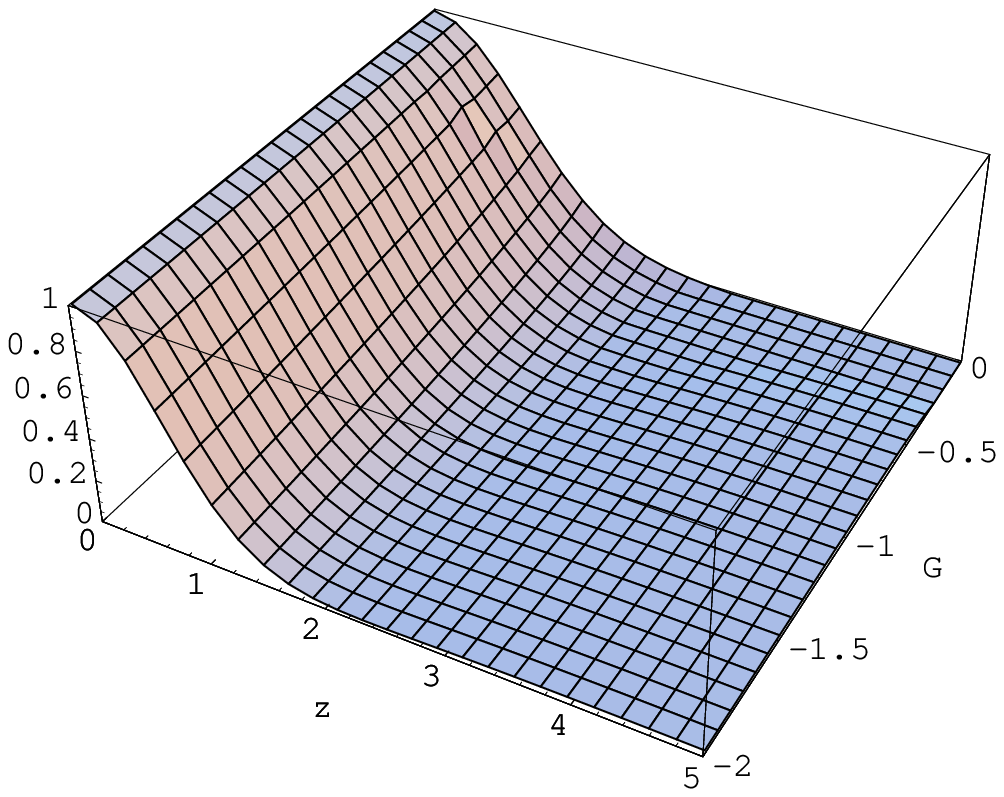}
\hspace{5mm}
\includegraphics*[angle=0,width=0.42\textwidth]{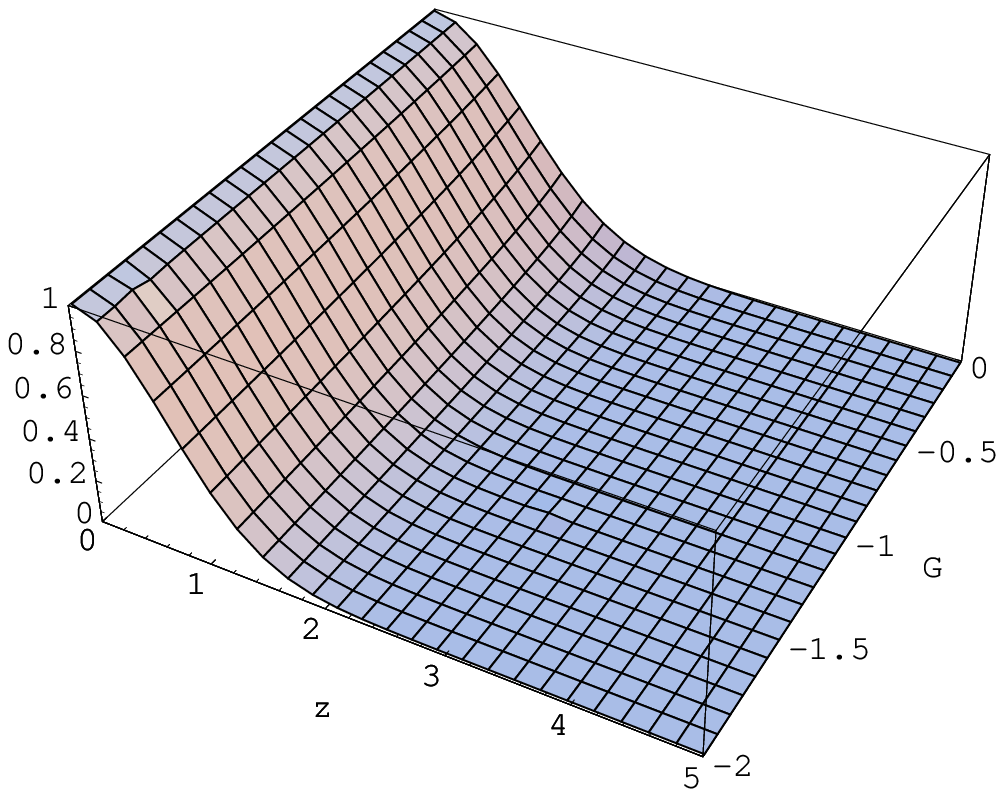}
\end{center}
\vspace{0.5mm}\caption{Subdiagonal Pad\'e approximants $[3,4]$, $g=\xi=1$,  for the series (\ref{20}) (left plot) and (\ref{23}) (right plot). It is seen that both approximations mimic the exact result (\ref{24}) very well. It may be also checked that the results weakly depend on the order of approximation.}
\end{figure}

\end{document}